\input{aipcheck}

\documentclass[
    ,final            
  ]
  {aipproc}

\layoutstyle{6x9}

\begin{document}

\title{Kaon Decays from AdS/QCD \footnote{Based on work done in collaboration with Thomas Hambye, Babiker Hassanain and John March-Russell \cite{Hambye:2005up,Hambye:2006av}.}}


\author{Martin Schvellinger}{
  address={IFLP-CCT-La Plata, CONICET and \\ Departamento de F\'{\i}sica,
Universidad Nacional de La Plata\\
CC 67, (1900)  La Plata, Argentina} }

\begin{abstract} We briefly review one of the current applications of the AdS/CFT correspondence known as AdS/QCD and discuss about the calculation of four-point quark-flavour current correlation functions and their applications to the calculation of observables related to neutral kaon decays and neutral kaon mixing processes.

\end{abstract}

\maketitle


\section{Introduction}

During the last ten years the development of ideas derived from Maldacena's conjecture \cite{Maldacena:1997re} has been outstanding. The concrete ansatz relating the quantum field theory generating functional and the supergravity action was proposed in \cite{ansatz}. A key point is that this correspondence works as a strong/weak coupling duality, in the sense that the strongly coupled regime of gauge theories can be described through the weak coupling regime of string theory. This fact arose interest in applying the gauge/gravity duality to construct a supergravity dual description of the strongly coupled regime of QCD.

The initial example worked out by Maldacena was the large $N$ limit of $SU(N)$ ${\cal {N}}=4$ supersymmetric Yang Mills theory, that is dual to type IIB supergravity on AdS$_5$ $\times$ S$^5$, with $N$ units of the $F_5$-form flux through the five-sphere and a constant dilaton \cite{Maldacena:1997re}. This background is the near horizon limit of $N$ parallel D3-branes. Then, the obvious steps were to break both supersymmetry and conformal invariance. For instance, if the $S^5$ is replaced by the orbifold $S^5/Z_2$ the dual gauge theory preserves ${\cal {N}}=2$ supersymmetries \cite{Kachru:1998ys}. One can go further and consider placing $N$ D3-branes at the tip of the conifold. Its near horizon geometry is AdS$_5$ $\times$ T$^{1,1}$ and the dual gauge theory now is an $SU(N) \times SU(N)$ ${\cal {N}}=1$ supersymmetric Yang Mills theory with bifundamental fields \cite{Klebanov:1998hh}. By introducing $M$ fractional D3-branes on this geometry it is possible to induce a logarithmic RG-flow in the gauge theory whose gauge group becomes  $SU(N) \times SU(N+M)$. However, the background is singular at the IR \cite{Klebanov:2000nc}. The solution to this problem is achieved by replacing the conifold by the deformed conifold, leading to a regular solution everywhere which also presents a cascade of Seiberg dualities \cite{Klebanov:2000hb}. One can also deform the Klebanov-Strassler solution by introducing a gaugino mass term in the field theory, leading to a solution where supersymmetry is completely broken \cite{Kuperstein:2003yt}. It is interesting to mention that as in the case of the supersymmetric background \cite{Gubser:2004qj}, the non-supersymmetric background has a massless pseudo-scalar glueball in the supergravity fluctuation spectrum which is interpreted as the Goldstone boson of the spontaneously broken $U(1)_B$ baryon number symmetry \cite{Schvellinger:2004am}.

There are also other brane constructions leading to supergravity duals preserving ${\cal {N}}=1$ supersymmetries in 4d \cite{MN}, and also breaking supersymmetry completely. Among the non-supersymmetric models there are several explicit brane constructions \cite{Kruczenski}, being the Sakai-Sugimoto model \cite{Sakai} the most interesting one since it accounts for chiral symmetry breaking and has a very nice geometrical realisation of it.

The models described above are generically known as top-down approaches. The ultimate aim is to holographically describe QCD from strings, i.e. somehow to reproduce results from chiral perturbation theory ($\chi$PT) at low momentum and OPE properties at large momentum. But this is very hard indeed. On the other hand, there are more phenomenological approaches which also take advantage of the AdS/CFT correspondence \cite{Erlich,DaRold}. These approaches are much more simplistic and much less ambitious and, in principle, they do not assume that the low dimensional models can be uplifted to a fundamental string theory dual of QCD. Henceforth we will focus upon a particular minimal version of a bottom-up approach called AdS/QCD or holographic QCD.

The basic idea is to use a 5d model inspired in the AdS/CFT correspondence to describe certain properties of the neutral kaon decays and kaon mixing processes. The material presented here is described in detail in the papers \cite{Hambye:2005up} and \cite{Hambye:2006av}. The 5d model is a reduced version of the so-called AdS/QCD proposed in \cite{Erlich,DaRold}. For reduced model we mean that we do not include the 5d scalar field and, therefore, the resulting version of the model has not the ability to describe the chiral condensate. The reason to turn off that field is our interest in exploring the predictions of the minimal model to obtain four-point functions. The full model would introduce further complications to the actual calculations that we want to avoid, since {\it a posteriori} we confirmed that the relevant physics is well described in this minimal setup.

Concretely, we shall describe the $\Delta I=1/2$ selection rule and the $B_K$ parameter through 5d calculations of four-point correlators. We regard our results as an additional non-trivial test for AdS/QCD since we consider calculations of four-point functions using interactions in the 5d gauge theory. In addition, a reasonably good comparison to experimental data (within $25 \%$ or better) has been obtained.

\section{A minimal AdS/QCD model}

As mentioned, there is a complementary approach to QCD using holographic dual models. In particular, we shall introduce an effective 5d theory with the basic elements to holographically describe certain properties of QCD \cite{Erlich,DaRold}\footnote{See also \cite{Hirn}.}. Let us consider the 5d model in some detail. The AdS$_5$ spacetime metric is ($L_0 \leq z \leq L_1$)
\begin{equation}
ds^2 = a^2(z) \, (\eta_{\mu\nu} \, dx^\mu dx^\nu - dz^2)
\, \ \,\,\,\,\,\,\,\,\,\,\, a(z)=L/z \ .
\end{equation}
We are only interested in 4d QCD quark-flavour current operators such as
\begin{eqnarray}
&{\hat{j}}^{\mu, \, a}_L \equiv {\bar q}_L \gamma^\mu t^a q_L \ ,& \\[.5em]
&{\hat{j}}^{\mu, \, a}_R \equiv {\bar q}_R \gamma^\mu t^a q_R \ ,&
\end{eqnarray}
where $q$ can be $u$, $d$ and $s$. We work in the chiral limit of QCD, therefore global flavour currents are conserved and the 4d quantum field theory global symmetry group is $SU(3)_L \times SU(3)_R$. Therefore $SU(3)_L \times SU(3)_R$ gauge fields are expected in the bulk, $L_{\mu}^a(x, z)$ and $R_{\mu}^a(x, z)$. The AdS/QCD ansatz for the 4d quantum field theory generating functional in terms of the dual 5d theory is given by
\begin{eqnarray}
&& \left\langle\exp \left( \displaystyle\int_{b} d^4x\, \left[ \hat{j}^{\mu,
\, a}_L(x)\, L_{\mu}^a(x, 0) + \hat{j}^{\mu, \, a}_R(x)\, R_{\mu}^a(x, 0)
\right] \right) \right\rangle_{QCD} \quad\qquad\qquad\qquad\qquad \nonumber \\[.5em]
&& \qquad\qquad\qquad\qquad\qquad\qquad\quad\qquad \approx ~~
\exp \left( - \displaystyle\int d^4x\,
dz\; {\cal {L}}_{5d}\bigg|_{V_\mu=v_\mu \cdots} \right) \ .
\end{eqnarray}
Then, for $n$-point quark-current correlators we just need to functionally derive $n$-times both sides in the above ansatz. The bulk theory is a gauged $SU(3)_L \times SU(3)_R$ 5d theory. The Lagrangian is
\begin{eqnarray}
{\cal {L}}_{5d} = \sqrt{g} \, M_5 \, Tr \left( -\frac{1}{4} \, L_{MN} \, L^{MN} - \frac{1}{4} \, R_{MN} \, R^{M} \right)\ ,
\end{eqnarray}
where  $M=(\mu, 5)$,  and $\mu=1, \cdot \cdot \cdot, 4$. The AdS/CFT relation between the dimension of the boundary theory operators and masses of the 5d fields is $(\Delta + 1) (\Delta - 3)=m_5^2$. Therefore, if $\Delta=3$, then $m_5=0$. Also, we have used the following definitions for field strengths $L_{MN}=\partial_ML_N-\partial_NL_M-i[L_M,L_N]$ and
$R_{MN}=\partial_MR_N-\partial_NR_M-i[R_M,R_N]$. Now, it is useful to define the following transformation of the 5d fields such that $V_M = (L_M + R_M)/\sqrt{2}$ and $A_M = (L_M - R_M)/\sqrt{2}$. To eliminate the mixing between $V_\mu$ and $V_5$ components, and between $A_\mu$ and $A_5$ components, we include $R_\xi$ gauge fixing terms as follows
\begin{eqnarray}
{\cal {L}}^V_{GF} &=&  - \frac{M_5 \, a}{2 \xi_V} \, Tr \left(\partial_\mu V_\mu - \frac{\xi_V}{a} \, \partial_5(a V_5) \right) \ ,  \\[.6em]
{\cal {L}}^A_{GF} &=&  - \frac{M_5 \, a}{2 \xi_A} \, Tr \left(\partial_\mu A_\mu - \frac{\xi_A}{a} \, \partial_5(a A_5) \right) \ ,
\end{eqnarray}
where $\xi_{V,A}$ are taken to be infinity.

As mentioned we work in the chiral limit, so that there is no explicit chiral symmetry breaking. On the other hand, chiral symmetry is spontaneously broken by imposing the following boundary conditions on the 5d gauge fields: $V_\mu|_{UV}=v_\mu$ and $\partial_z V_\mu|_{IR}=0$, and  $A_\mu|_{UV}=a_\mu$ and $A_\mu|_{IR}=0$, which imply that there are not zero modes for these fields. In addition, we impose $\partial_z (A_5/z)|_{UV}=\partial_z (A_5/z)|_{IR} =0$, leading to a zero mode for $A_5$.

The 5d propagators have been obtained in \cite{Hambye:2005up,Hambye:2006av}. The vector and axial-vector bulk-to-bulk propagator have transverse and longitudinal parts. We work in a gauge such that $V_5=0$. There are also bulk-to-bulk and bulk-to-boundary $A_5$ propagators. The bulk-to-boundary vector and axial-vector propagators are transversal because the corresponding 4d global currents at the boundary are conserved. The two-point axial-vector current correlator in the limit $p\to 0$ is $\Pi_A(p^2)\vert_{p=0}= F_{\pi}^2 =2M_5L/(L_1^2-L_0^2)$.

Two-point quark current correlators can easily be calculated within the 5d holographic dual model. In the large $N$ limit they are given by the following expressions
\begin{eqnarray}
\Pi_V(p^2) &=& \sum_n\frac{F^2_{V_n}}{p^2-M^2_{V_n}} \ , \\[.6em]
\Pi_A(p^2) &=& \sum_n\frac{F^2_{A_n}}{p^2-M^2_{A_n}} + F^2_\pi \ .
\end{eqnarray}
The ratio of lowest-lying meson masses from the model $M_{a_1}/M_\rho|_{AdS} = 1.6$ agrees well with the experimental ratio $M_{a_1}/{M_\rho}|_{Experimental} = 1.63$. Decay constants of the lowest-lying mesons agree very well with experiments \cite{Erlich,DaRold,Hirn}.

Next step is to calculate the $n$-point quark-flavour current correlators, since it allows to study the effect of interactions in the bulk. In particular, four-point correlators have a very interesting application to QCD related to kaon decays and mixing processes. So, we shall show how to calculate the $\Delta I=1/2$ selection rule and $B_K$ parameter using this holographic 5d dual model \cite{Hambye:2005up,Hambye:2006av}.

\section{The $\Delta I=1/2$ rule and the $B_K$ parameter}

When CP violation effects are neglected, two independent $K \rightarrow \pi \pi$ decay channels ($K^0 \rightarrow \pi^+ \pi^-$ and $K^0 \rightarrow \pi^0 \pi^0$) are present. The amplitudes corresponding to these decay channels can be written in terms of the $\Delta I=1/2$ amplitude $A_0$ and the $\Delta I = 3/2$ amplitude $A_2$ in the following form
\begin{eqnarray}
A(K^0 \rightarrow \pi^+ \pi^-) &=& A_0 \,\, e^{i \delta^0} + \sqrt{1/2} \,\, A_2 \,\, e^{i \delta_2} \ , \nonumber
\\[.5em]
A(K^0 \rightarrow \pi^0 \pi^0) &=& A_0 \,\, e^{i \delta^0} - \sqrt{2} \,\, A_2 \,\, e^{i \delta_2} \ . \nonumber
\end{eqnarray}
The experimental values are $\mbox{Re}A_0=2.72 \cdot 10^{-4}~{\mbox{MeV}}$ and
${\mbox{Re}A_2}=1.22 \cdot 10^{-5}~{\mbox{MeV}}$. This leads to
\begin{eqnarray}
\frac{1}{\omega}\, & \equiv & \, \frac{\mbox{Re}A_0}{\mbox{Re}A_2}\,\equiv\, \frac{\mbox{Re}(K \rightarrow(\pi\pi)_{I=0})}{\mbox{Re}(K\rightarrow (\pi \pi)_{I=2})}\,=\,22.2 \ ,
\end{eqnarray}
and the large value of Re$A_0/$Re$A_2$ is the so called $\Delta I = 1/2$ rule.

We use AdS/QCD to calculate $\mbox{Re}A_0$ and $\mbox{Re}A_2$ in the chiral limit. In this limit, at order $p^2$, all $\Delta S = 1$ transitions can be obtained from a $\Delta S = 1$ effective Lagrangian involving the octet and 27 coupling constants \cite{Hambye:2003cy,KMW}
\begin{eqnarray}
{\cal {L}}_{eff}^{\Delta S=1}=-\frac{G_F}{\sqrt{2}}V_{ud}V_{us}^*\left[ g_{8}\;
{\cal {L}}_{8} + g_{27}\;{\cal {L}}_{27} \right] \ ,
\end{eqnarray}
\vskip-2mm
\begin{eqnarray}
{\cal {L}}_8 = \displaystyle{\sum_{i=1,2,3}({\cal {L}}_{\mu})_{2i}\
({\cal {L}}^{\mu})_{i3}} \ ,
\end{eqnarray}
\vskip-2mm
\begin{eqnarray}
{\cal {L}}_{27} = \frac{2}{3}({\cal {L}}_{\mu})_{21}\ ({\cal {L}}^{\mu})_{13}
+ ({\cal {L}}_{\mu})_{23}\ ({\cal {L}}^{\mu})_{11} \ ,
\end{eqnarray}
where
\begin{equation}
{\cal {L}}_{\mu} = - i F_{\pi}^2\ U(x)^{\dagger} D_{\mu} U(x) \ ,
\end{equation}
and $V_{ud}=0.974$, $V_{us}=0.224$. The octet term proportional to $g_{8}$ induces pure $\Delta I=1/2$ transitions, while the term proportional to $g_{27}$ induces both $\Delta I=1/2$ and $\Delta I=3/2$ transitions. In this way $1/\omega$ becomes
\begin{eqnarray}
 \frac{1}{\omega} &=& \frac{\mbox{Re}A_0}{\mbox{Re}A_2}  =   \frac{9}{5\sqrt{2}} \left(\frac{g_{8}+\frac{1}{9} g_{27}}{g_{27}} \right) \ .
\end{eqnarray}
 Now, we need to calculate $g_8$ and $g_{27}$. To do it we separate the long and short distance contributions and perform an OPE, obtaining the effective Hamiltonian for $|\Delta S|=1$ transitions \cite{GLAM},
\begin{eqnarray}
{\cal {H}}_{ef\hspace{-0.5mm}f}^{\Delta S=1}=\frac{G_F}{\sqrt{2}}\;
\xi_u\sum_{i=1}^8 c_i(\mu)Q_i(\mu)\ ,
\hspace{1cm}(\mu <m_c=\textrm{charm quark mass}) \ ,
\end{eqnarray}
where
\begin{equation}
c_i(\mu)=z_i(\mu)+\tau\, y_i(\mu) \ ,
\end{equation}
where $\tau=-\xi_t/\xi_u$ and $\xi_q=V_{qs}^*V_{qd}^{}$. The arbitrary renormalisation scale $\mu$ separates short and long distance contributions to the decay amplitudes. The local four-quark operators $Q_i(\mu)$ after Fierz reordering can be written in terms of color singlet quark bilinears: (sum over $q=u,d,s$ is understood)
\begin{eqnarray}
Q_1 &=& 4\,\bar{s}_L\gamma^\mu d_L\,\,\bar{u}_L\gamma_\mu u_L \ ,
\,\,\,\,\,\,\,\,\,\,\,\,\,\,\,\,\,\,\,\,\,\,\,\,\,\,\,\,\,\,\,\,\,\,\,\,
Q_2 \,\,\,\,\, = \,\,\,\,\, 4\,\bar{s}_L\gamma^\mu u_L\,\,\bar{u}_L
\gamma_\mu d_L\ ,  \nonumber \\[.75em]
Q_3 &=& 4\,\sum_q \bar{s}_L\gamma^\mu d_L\,\,\bar{q}_L\gamma_\mu q_L \ ,
\,\,\,\,\,\,\,\,\,\,\,\,\,\,\,\,\,\,\,\,\,\,\,\,\,
Q_4 \,\,\,\,\, = \,\,\,\,\, 4\,\sum_q \bar{s}_L\gamma^\mu q_L\,\,\bar{q}_L
\gamma_\mu d_L\ ,  \nonumber \\[.5em]
Q_5 &=& 4\,\sum_q \bar{s}_L\gamma^\mu d_L\,\,\bar{q}_R\gamma_\mu q_R \ ,
\,\,\,\,\,\,\,\,\,\,\,\,\,\,\,\,\,\,\,\,\,\,\,\,
Q_6 \,\,\,\,\, = \,\,\,\,\, -8\,\sum_q \bar{s}_L q_R\,\,\bar{q}_R
d_L\ ,  \nonumber \\[.5em]
Q_7 &=& \,4\,\sum_q \frac{3}{2}e_q\,\bar{s}_L\gamma^\mu d_L\,\,
\bar{q}_R \gamma_\mu q_R \ , \,\,\,\,\,\,\,\,
Q_8 \,\,\,\,\, = \,\,\,\,\, -8\,\sum_q \frac{3}{2}e_q\,\bar{s}_L q_R\,\,
\bar{q}_R d_L \ ,
\end{eqnarray}
with the definitions $q_{R,L} = \frac{1}{2}(1\pm\gamma_5)\, q$ and $e_q = (2/3,\,-1/3,\,-1/3)$.
\medskip

Now, in order to calculate the non-factorisable contribution to $g_8$ and $g_{27}$ we use the chiral symmetry properties of the $\Delta S =1$ effective Lagrangian. Instead of calculating the $K \rightarrow \pi\pi$ amplitude explicitly, it is much simpler to calculate them taking $U=1$, i.e. considering the processes with no external pseudoscalar and only two external sources coming from the covariant derivatives of $U$. It is convenient to consider the processes with two external right-handed sources. The non-factorisable contribution to this process of four-quark operators is then given by Green's functions involving the two left-handed currents of the four-quark operator inducing this process and the two right-handed currents coupling to the two right-handed sources.

Including the leading $N_c$ non-factorisable contribution from $Q_{1}$ and $Q_2$, $g_8$ and $g_{27}$ are given by the $Q^2$ integrals (with $Q$ the Euclidean momentum flowing between the two left-handed currents) of the two Green's functions \cite{Hambye:2003cy,Peris}:
\begin{eqnarray}
g_8 (\mu) = z_1 \biggl(-1+\frac{3}{5}g_{\Delta S = 2} \biggr) +
z_2 \biggl(1-\frac{2}{5}\, g_{\Delta S = 2} - \int^{\mu^2}_{0} dQ^2\;
\frac{{W}_{LLRR}(Q^2)}{4 \pi^2 F_\pi^2}\biggr) \ ,
\end{eqnarray}
\begin{eqnarray}
g_{27}(\mu)=\frac{3}{5}\, (z_1+z_2)\, g_{\Delta S = 2}\ ,
\end{eqnarray}
\begin{eqnarray}
g_{\Delta S = 2}(\mu) = 1-\frac{1}{32\pi^2 F_\pi^2}\, \int^{\mu^2}_{0}
dQ^2\; W_{LRLR}(Q^2) \ ,
\end{eqnarray}
with
\begin{eqnarray}
{W}_{LRLR}(Q^2)&=&- \frac{4}{3} \frac{Q^2}{F_\pi^2}\eta_{\alpha\beta}\eta_{\mu\nu} \int \frac{d\Omega_{q}}{4 \pi}\; {W}_{LRLR}^{\mu\alpha\nu\beta}(q) \ , \\[.5em]
{W}_{LLRR}(Q^2)&=&-\frac{1}{3} \frac{Q^2}{F_\pi^2} \eta_{\alpha\beta} \eta_{\mu\nu} \int \frac{d\Omega_{q}}{4 \pi}\; {W}_{LLRR}^{\,\mu\,\nu\,\alpha\,\beta}(q)\ ,
\end{eqnarray}
where
\begin{eqnarray}
&&\!\!\!\!\!\!\!\!\!\!\!\!\!\!\!
{W}_{LRLR}^{\mu\alpha\nu\beta}(q) =
\lim_{l\rightarrow 0}\ i^3\!\!\! \int d^4x\, d^4y\, d^4z\; e^{iqx+il(y-z)}
\langle 0\vert {\hat T} \{ L_{\bar{s}d}^{\mu}(x)R_{\bar{d}s}^{\alpha}(y)
L_{\bar{s}d}^{\nu}(0)R_{\bar{d}s}^{\beta}(z)\}
\vert 0\rangle\vert_{\mbox{\rm\tiny conn}} \ , \nonumber \\&&\\[.5em]
&&\!\!\!\!\!\!\!\!\!\!\!\!\!\!\!
{W}_{LLRR}^{\mu\nu\alpha\beta}(q) =
\lim_{k\rightarrow 0} \ i^3\!\!\!\int d^4x d^4y d^4z\ e^{iqx + ik(y-z)} \langle 0\vert {\hat T} \{L_{\bar{s}u}^{\mu}(x)L_{\bar{u}d}^{\nu}(0) R_{\bar{d}u}^{\alpha}(y)R_{\bar{u}s}^{\beta}(z)\}\vert
0\rangle\vert_{\mbox{\rm\tiny conn}} \ .\nonumber\\
\end{eqnarray}
In order to calculate the integrals above the $Q^2$ dependence of $W_{LLRR}$ and $W_{LRLR}$ must be determined. However, this dependence is well-known only in both asymptotic regimes, i.e. $Q^2 \to 0$ and $Q^2 \to \infty$ limits. In the low momentum limit, from $\chi$PT \cite{Peris,BP1}
\begin{eqnarray}
W_{LRLR}(Q^2)&=&6-24 \left( {2 l_1+ 5 l_2 + l_3 + l_9} \right)\;
\frac{Q^2}{F_\pi^2} + \ldots \label{chiralLRLR} \ ,  \\[.5em]
W_{LLRR}(Q^2)&=&-\frac{3}{8} + \left( -\frac{15}{2}l_{3}+\frac{3}{2}l_9
\right)\; \frac{Q^2}{F_{\pi}^2} + \ldots \label{chiralLLRR} \ ,
\end{eqnarray}
where the $l_i$ are the standard chiral Lagrangian coefficients. In the large $N_c$ limit and using Witten diagrams, one can see that $W_{LRLR}$ and $W_{LLRR}$ are given by a sum of simple to triple poles in $Q^2$ multiplied by polynomials in $Q^2$. Using this, the most general form for the Green's functions is given by \cite{Hambye:2003cy,Peris}:
\begin{eqnarray}
 W_{LRLR}&=&
\sum_{i=1}^{\infty}\left(\frac{\alpha_i}{(Q^2+M^2_{i})}+\frac{\beta_i} {(Q^2+M^2_{i})^2}+\frac{\gamma_i}{(Q^2+M^2_{i})^3}\right) ~,\label{generalLRLR}  \\[.5em]
 W_{LLRR}&=& \sum_{i=1}^{\infty}\left(\frac{\alpha'_i}{(Q^2+M^2_{i})}+\frac{\beta'_i} {(Q^2+M^2_{i})^2}+\frac{\gamma'_i}{(Q^2+M^2_{i})^3}\right) \ . \label{generalLLRR}
\end{eqnarray}

In addition, there is another important related observable, the $B_K$ parameter, which parameterizes the $K^0-\bar{K}^0$ mixing. At the quark level, $K^0$ and $\bar{K}^0$ mix due to a box one-loop diagram where the $K^0$ transforms itself into $\bar{K}^0$ through a pair of W bosons. This diagram leads to the following effective Hamiltonian \cite{Herrlich}:
\begin{eqnarray}
{\cal {H}}_{eff}^{\Delta S=2} = \frac{G_{F}^{2}M_{W}^2}{4\pi^2}\,
\left[\lambda_{c}^2F_1+\lambda_{t}^2F_2+2\lambda_{c}\lambda_{t}F_3\right]
\times C_{\Delta S=2}(\mu)\;
Q_{\Delta S=2}(x) \ ,
\end{eqnarray}
where
\begin{equation}
Q_{\Delta S=2}(x) \equiv (\bar{s}_{L}(x)\gamma^{\mu}d_{L}(x))\,
(\bar{s}_{L}(x)\gamma_{\mu}d_{L}(x))
\end{equation}
and $C_{\Delta S = 2}$ is the Wilson coefficient. From this effective Hamiltonian, one can define the matrix element
\begin{equation}
\langle\bar{K}^0|Q_{\Delta S=2}(0)|K^0\rangle \equiv \frac 4 3\,
f_{K}^2\, M_{K}^2\, B_{K}(\mu) \ ,
\end{equation}
and then
\begin{equation}
\hat{B}_K \equiv C_{\Delta S = 2}(\mu)\, B_K(\mu) \ .
\end{equation}
The large $N_c$ limit (i.e. the factorisable contribution) gives $B_K=3/4$. In the chiral limit and at leading $N_c$ order, it turns out that the non-factorisable contribution is determined by the same integral of $W_{LRLR}$
\begin{equation}
B_K(\mu)= \frac{3}{4}\, g_{\Delta S=2}(\mu) \ .
\end{equation}

\section{Holographic calculation of 4-point functions}

The results of this section are presented in \cite{Hambye:2005up,Hambye:2006av}. There are three classes of 5d holographic  Witten diagrams that contribute to a general four-current correlator in momentum space:  diagrams where $A_5$ propagates, X-diagrams involving the four-boson vertex, and Y-diagrams, which involve two three-boson vertices. Including all the contributions, we find that the two correlators are proportional to each other
\begin{eqnarray}
& W_{LRLR}(Q^2) = \displaystyle{\frac{4i}{3}\frac{Q^2}{F_\pi^2}\Sigma(p=iQ)}
\ , & \\[.5em]
& W_{LLRR}(Q^2) = -\displaystyle{\frac{i}{12}\frac{Q^2}{F_\pi^2}\Sigma(p=iQ)} \ , &
\end{eqnarray}
where the sum of the diagrams is given by $\Sigma=\Sigma_{X}+\Sigma_{A_5}+\Sigma_{Y}$. The expressions for the individual $\Sigma$ contributions are given in \cite{Hambye:2006av} in terms of integrals in $z$. The integrals can all be done analytically. We perform all the integrals with the limits $L_1$ and $L_0$, then take the limit $L_0\to 0$. All the divergent contributions cancel, and we obtain
\begin{eqnarray}
&\Sigma(Q)_{L_0 \to 0} = \displaystyle{3iM_5L\left[\frac{16}{Q^6L_1^6} -
\frac{14}{5Q^4L_1^4}-\frac{299}{240I_1^2}+ \frac{7}{20Q^2L_1^2I_1^2} +
\frac{299}{240I_0^2}-\frac{2}{15Q^2L_1^2I_0^2} \right.} & \nonumber \\[.5em]
{} &  \left. +\,\displaystyle{\frac{7}{5Q^4L_1^4I_0^2} +
\frac{16}{Q^6L_1^6I_0^2}-\frac{32}{Q^6L_1^6I_0} +
\frac{13}{12QL_1I_0I_1}} \right] \ , &
\end{eqnarray}
where $I_{0,1}=I_{0,1}(QL_1)$ are the modified Bessel functions of zeroth and first order, respectively.
Note that $W_{LRLR}$ is found to be positive definite, while $W_{LLRR}$ is negative definite, as expected from $\chi$PT. Both correlators also approach zero as $Q \to \infty$, and satisfy the ``sum of poles'' functional form.

The pole structure of the propagators for low momentum constitutes a strong check on our calculation. As explained above, $\chi$PT gives us a constraint on the behaviour of the correlators as $Q\to 0$. Taking that limit in the expression of $\Sigma(Q)$, we obtain
\begin{eqnarray}
\lim_{Q^2\rightarrow 0}\Sigma(Q)&=&3iM_5L\left(\frac{-3}{Q^2L_1^2} + \frac{105}{64} - \frac{1521}{2560}Q^2L_1^2 + {\cal {O}}(Q^4)\right) \ .
\end{eqnarray}
This is the functional form required by $\chi$PT. The $Q^2$ pole is due to the massless pions. Therefore, our correlators have the low $Q$ behaviour given by
\begin{eqnarray}
\lim_{Q^2\rightarrow 0}W_{LRLR}(Q^2) &=& 6-\frac{105M_5L}{16}\;
\frac{Q^2}{F_\pi^2} +{\cal {O}}(Q^4) \ , \\[.5em]
\lim_{Q^2\rightarrow 0}W_{LLRR}(Q^2) &=& -\frac{3}{8}+\frac{105M_5L}{256}\;
\frac{Q^2}{F_\pi^2} +{\cal {O}}(Q^4) \ .
\end{eqnarray}
By comparison with the expressions obtained using $\chi$PT in the chiral limit, we found that there is a very good matching for $W_{LRLR}$, for the range of validity of $\chi$PT. On the other hand, $W_{LLRR}$ does not exhibit a good match with the $\chi$PT calculations. Since in \cite{Hirn} the chiral Lagrangian coefficients were calculated in the holographic dual setting, our predictions for low momentum can be compared with those of $\chi$PT with the AdS $l_i$ coefficients calculated in \cite{Hirn}. We have done this to understand the discrepancy in $W_{LLRR}$. We also have done a variety of consistency checks and we do not see any possibility of deviations which would alter the proportionality between $W_{LLRR}$ and $W_{LRLR}$. This makes us confident that our results are correct. It seems possible to us that the problem might lie with the sole and rather subtle $\chi$PT calculation of the $l_3$ dependence of $W_{LLRR}$. Note that this difference for the $l_3$ coefficient for $W_{LLRR}$ is not large enough to alter the fact that below we find a large enhancement for $g_8$.

\section{Results and Conclusions}

The results are shown in Table 1. Columns $A, B$ show a fitting to the rho meson mass $m_{\rho}$, the $a_1$ meson mass $m_{a_1}$, pion decay constant $F_{\pi}$, $g_8$ and $g_{27}$, while $1/\omega$ and $\hat{B}_K$ are the model predictions. Columns $C, D$ show a fitting to $m_{\rho}$, $m_{a_1}$, $F_{\pi}$, $1/\omega$ and $\hat{B}_K$, being $g_8$ and $g_{27}$ the predictions of AdS/QCD. Note that, as explained in section 4 of \cite{Hambye:2006av}, $g_{8}^{TOT}=g_8+1.8$, $g_{27}^{TOT}=g_{27}+0.06$, where $g_8$ and $g_{27}$ are the quantities we calculated from the AdS model. We use the values $F_{\pi}=87$ MeV, $m_{\rho}^{exp}=776$ MeV, $m_{a_1}^{exp}=1230$ MeV, $g_{8}^{exp}=5.1$ and $g_{27}^{exp}=0.29$. The UV energy cutoffs that we used are 1300 MeV and 1500 MeV. We have used these values to show that our calculations are not sensible to changes of the UV limit of integration. We have chosen these specific values since for them there are values for the Wilson coefficient in the literature (see references in \cite{Hambye:2006av}).
\vskip1mm

\begin{table}[h]
\caption{Fittings and predictions of AdS/QCD for a number of observables. Details are described in the text.}
\begin{tabular}{ccccc}
~ & ~ \\[.3em]
\hline
   \tablehead{1}{c}{b}{Observable}
  & \tablehead{1}{c}{b}{A}
  & \tablehead{1}{c}{b}{B}
  & \tablehead{1}{c}{b}{C}
  & \tablehead{1}{c}{b}{D}\\[.1em]
\hline
$L_0^{-1}$                    & $1300$ MeV & $1500$ MeV & $1300$ MeV & $1500$ MeV \\[.4em]
$L_1^{-1}$                    & $274$  MeV & $275$  MeV & $277$  MeV & $280$  MeV \\[.4em]
$m_\rho^{th}/m_\rho^{exp}$    & $0.91$     & $0.90$     & $0.93$     & $0.92$     \\[.4em]
$m_{a_1}^{th}/m_{a_1}^{exp}$  & $0.95$     & $0.93$     & $0.97$     & $0.95$     \\[.4em]
$F_\pi^{th}/F_\pi$            & $1.15$     & $1.17$     & $1.12$     & $1.14$     \\[.4em]
$g_{8}^{TOT}/g_{8}^{exp}$     & $0.74\tablenote{fitted}$     & $0.72^*$     & $0.75^\dagger$   & $0.74^\dagger$   \\[.4em]
$g_{27}^{TOT}/g_{27}^{exp}$   & $0.85^*$     & $0.85^*$     & $0.79^\dagger$   & $0.78^\dagger$   \\[.4em]
$1/\omega$                    & $19.5$\tablenote{predicted}   & $19.2^\dagger$   & $21.4^*$     & $21.3^*$     \\[.4em]
$\hat{B}_K^{th}$              & $0.38^\dagger$   & $0.38^\dagger$   & $0.34^*$     & $0.34^*$     \\[.1em]
\hline
\end{tabular}
\label{tab:a}
\end{table}

Our results are similar to those obtained in other analytical calculations using the $1/N_c$ expansion \cite{Hambye:2003cy,Hambye}. However, the two methods are quite different, being the main advantage of our model the fact that it allows to calculate the four-point functions in the entire momentum range within one consistent setting. This is very interesting since it avoids the need for interpolation in any specific momentum range.

Another interesting feature of AdS/QCD is that it includes the contributions of the infinite tower of meson resonances to the four-point correlators in a consistent and automatic way.  The holographic calculation reproduces to a good level of accuracy the low-momentum and high-momentum behaviour of these correlators as deduced from $\chi$PT and perturbative QCD, respectively. The exchange of meson resonances modifies the momentum dependence at the intermediate regime. Indeed, an impressively good agreement is obtained in the low and high momentum limit for the correlator $W_{LRLR}$. In addition, the results of a fit of the holographic predictions agree well with the experimental data. This shows that the dynamics of the $\Delta I= 1/2$ rule is operative in AdS/QCD. For quantities as difficult to calculate as the isospin amplitudes Re$A_0$ and Re$A_2$ this is remarkable.

A limitation of the model concerns the description of $\chi$SB. Although the IR boundary conditions on the bulk $SU(3)_L \times SU(3)_R$ gauge fields that we use correctly incorporate the leading $\chi$SB behaviour, a bi-fundamental bulk scalar is needed to fully account for the physics of $\chi$SB.  The inclusion of this field will directly introduce pseudo-scalar resonances into the 4d field content, and we would expect these will have relevant contributions to the four-current correlators calculated in \cite{Hambye:2003cy,Hambye}. We will also have an extra parameter that can be tuned \cite{Erlich,DaRold}, corresponding to the quark condensate. We have also not included the effects of the anomalous $U(1)_A$ symmetry of QCD, nor the explicit breaking of chiral symmetry due to bare quark masses. One envisions these improvements having a complicated yet positive effect on the calculation of four-point current correlators presented in \cite{Hambye:2005up,Hambye:2006av}.

We have shown how AdS/QCD provides a method to calculate the contribution of four-point quark correlators at intermediates energies. In summary, there is a good agreement with experimental data within $25\%$ of accuracy or better, depending on the observable. Also, our calculations improve the understanding of the role of the interactions in the bulk by comparison with observables related to neutral kaon decays and kaon mixing processes. We think that the results of \cite{Hambye:2005up,Hambye:2006av} encourage further investigations on AdS/QCD including physical processes involving $n$-point correlation functions.


\begin{theacknowledgments}

I am indebted to Thomas Hambye, Babiker Hassanain and John March-Russell
for collaboration on the papers \cite{Hambye:2005up,Hambye:2006av}. I
acknowledge the Rudolf Peierls Centre for Theoretical Physics at the
University of Oxford where most of this work has been done. Also kind
hospitality at the ECT$^*$, Trento, as well as the Newton Institute,
Cambridge, and the Instituto de Astronom\'{\i}a y F\'{\i}sica del Espacio
(IAFE-UBA-CONICET) is acknowledged.

\end{theacknowledgments}



\bibliographystyle{aipproc}   

\end{document}